\documentclass[useAMS,usenatbib]{mn2e}
\usepackage{times,amsmath,amssymb,units,graphicx,txfonts,aas_macros,natbib}
\usepackage{ulem}

\renewcommand\emph[1]{\textit{#1}}

\newcommand\Pm{\mathrm{Pm}}
\newcommand\Rm{\mathrm{Rm}}
\newcommand\Co{\hat{\Omega}}

\newcommand\tms{\!\times\!}
\newcommand\cdt{\!\cdot\!}

\newcommand\xx{\hat{{\mathbf x}}}
\newcommand\yy{\hat{{\mathbf y}}}
\newcommand\zz{\hat{{\mathbf z}}}

\newcommand{\mn}[1]{\overline{#1}}

\newcommand\V{\mathbf{v}}
\newcommand\U{\mathbf{u}}
\newcommand\mU{\mn{\mathbf{u}}}
\newcommand\B{\mathbf{B}}
\newcommand\mB{\mn{\mathbf{B}}}
\newcommand\EMF{\mbox{\boldmath{${\cal E}$}}}
\newcommand\Tf{\mathcal{B}}
\newcommand\mTf{\bar{\mathcal{B}}}

\newcommand\aK{\alpha_{\rm kin}}
\newcommand\aM{\alpha_{\rm mag}}

\newcommand{\simgt}%
           {\,\hbox{\lower0.6ex\hbox{$\sim$}\llap{\raise0.6ex\hbox{$>$}}}\,}

\title[Mean-field approach to stratified MRI]%
      {A mean-field approach to the propagation of field patterns\\
       in stratified magneto rotational turbulence}
\author[O.~Gressel]%
       {Oliver~Gressel\thanks{E-mail:o.gressel@qmul.ac.uk}\\
        Astronomy Unit, Queen Mary, University of London, %
        Mile End Road, London E1 4NS, United Kingdom
       }

\begin{document}

\date{Accepted 1988 December 15. %
      Received 1988 December 14; %
      in original form 1988 October 11}

\pagerange{\pageref{firstpage}--\pageref{lastpage}} \pubyear{2002}

\maketitle

\label{firstpage}

\begin{abstract}

Local shearing box simulations of stratified magneto rotational
turbulence invariably exhibit cyclic field patterns which propagate
away from the disc midplane. A common explanation for this is magnetic
buoyancy. The recent analysis by Shi et~al. however shows that
the flow is buoyantly stable below one disc scale height $H$,
necessitating an alternative explanation in this region.

 We here conduct and analyse direct numerical
simulations to explain the observed behaviour by means of a mean-field
description. Apart from the mean radial and azimuthal field, we
monitor the small-scale current helicity, which we propose as a key
indicator for saturation.

Reconstructing the horizontally averaged field, we demonstrate that
the problem can be reduced to a one-dimensional induction equation.
By means of the so-called test field method, we then determine the
underlying closure parameters.   Our
analysis shows that, apart from a possible \emph{direct} MRI dynamo,
two distinct \emph{indirect} dynamo mechanisms operate in the
disc. This resolves the issue of the ``wrong'' sign of the MRI dynamo
effect.

Finally, we use the obtained closure parameters to run a dynamically
quenched dynamo model. This model approximately recovers the observed
field patterns in the mean fields.  Moreover,
the model reproduces the prevailing parity and the distinct phase
pattern in the small-scale current helicity. The latter property might
open a potential route to understand the saturation of MRI induced
turbulence.

\end{abstract}



\section{Introduction}

Recent numerical findings have challenged the sustainability of
hydromagnetic turbulence driven by the magneto rotational instability
\citep[MRI,][]{1998RvMP...70....1B}, and hence the viability of the
approach to provide the turbulent viscosity needed to explain
accretion luminosities \citep{2007MNRAS.376.1740K}.

The fundamental mechanism of the MRI relies on the interplay between
the epicyclic and magnetic restoring forces. A \emph{coherent}
magnetic field acts as the mediator to extract free energy from the
Keplerian rotation. Unless maintained by an external source or a
helical dynamo, such a coherent field is, however, prone to
dissipation by the turbulence created via parasitic instabilities
\citep{1994ApJ...432..213G,2009ApJ...698L..72P,2009MNRAS.394..715L}. This
becomes apparent when looking at the most simplified case:
	
\cite{2007ApJ...668L..51P} have analysed the results of a set of
unstratified local MRI simulations. The scaling law they derive
predicts turbulent stresses based on several input parameters. For the
case without vertical net-flux, they conclude that simulation results
should depend linearly on the numerical resolution.
\cite{2007A&A...476.1113F} have independently confirmed this
prediction and show that transport coefficients do not converge with
increasing resolution.  This finding of vanishing turbulent stresses
$W_{xy}$ for zero net flux (ZNF) is in contrast to the case of a net
vertical flux (NVF), which continuously drives long-wavelength MRI
modes \citep[see discussion in][]{2007ApJ...668L..51P}, and for which
convergence has been obtained \citep{2009arXiv0909.1570D}.

\cite{2007A&A...476.1113F} attribute the lack of convergence to a
resolution-dependent effective Prandtl number $\Pm_{\rm num}$ of their
code. In a subsequent paper, \cite{2007A&A...476.1123F} show that
convergence can be recaptured if explicit dissipation with $\Pm\simgt
2$ is included. For smaller values of $\Pm$, the lack of convergence
may be attributed to the fact that a small-scale dynamo becomes much
harder to excite \citep{2005ApJ...625L.115S}. An alternative
interpretation has been put forward by \cite{2009arXiv0911.5603K}, who
relate the convergence issue to the problem of resolving radial
structures occurring for non-axis symmetric MRI modes.

  But how realistic are such local models?
\cite{2009ApJ...696.1021V} has recently pointed out that unstratified
local simulations do not provide a (physically relevant) outer scale
for the turbulence, and therefore the observed $\Pm$ dependence might
be an artifact of the local approximation.

\cite{2009arXiv0909.1570D} indeed report characteristic differences in
magnetic power spectra for VNF and ZNF simulations, illustrating the
lack of a well-defined injection scale in the latter case. Since the
eddy viscosity is determined by the outer scale, this is sufficient to
explain the convergence in $W_{xy}$ for ideal MHD. Note that
\cite{2007MNRAS.378.1471L}, however, found a strong trend towards
weaker turbulence for smaller explicit $\Pm$ in unstratified NVF
simulations. 

Regardless of the $\Pm$ dependence, coherent field structures appear
to be the prerequisite for driving significant accretion
stresses. Moreover, large-scale fields may even contribute to the
accretion stresses directly \citep[see e.g. Fig.~8
in][]{2008A&A...490..501J}. In the absence of external fields, a
mean-field dynamo can provide a means to replenish coherent fields
that in turn drive MRI. This idea has first been advocated by
\cite{1995ApJ...446..741B}. Such an effect is likely beneficial to
sustain MRI turbulence, possibly even at $Pm\ll 1$ as the excitation
conditions of large-scale dynamos are known to be independent of $\Pm$
\citep{2009ApJ...697.1206B}. In fact, \cite{2009arXiv0909.1570D} have
very recently shown that stratification provides a sufficient
condition for convergence in the ZNF case and attribute this to a
dynamo-generated mean azimuthal field \citep[also
cf.][]{2009arXiv0909.2003S}. If this is the case, the saturation
amplitude of $W_{xy}$ should depend on the quenched state of the
underlying dynamo.

Along these lines, it becomes mandatory to better understand how
accretion discs can maintain coherent fields -- overcoming both
turbulent dissipation and buoyant field expulsion. As advocated in
\cite{2010AN....331..101B}, understanding the non-linear saturation of
MRI in a realistic scenario will require to couple a closure model for
turbulent stresses with mean-field dynamo theory. As for the former,
approaches beyond a simple Shakura-Sunyaev viscosity include
phenomenological closures by \cite{2003MNRAS.340..969O} (ZNF),
\cite{2006PhRvL..97v1103P} (VNF), and \cite{2008MNRAS.390..331H}. The
applicability of these models has, of course, to be checked by
comparison with sets of numerical simulations
\citep{2009AN....330...92L}.  Combining such closures with
(dynamically quenched) mean-field dynamo models will ultimately allow
to develop a consistent sub-grid scale framework, potentially enabling
global large-eddy simulations of ionised accretion discs.

In the following, we aim to demonstrate that there exist
helicity-based dynamo mechanisms in stratified MRI simulations which
are consistent with the observed rising field structures.


\section{Numerical setup}

We follow the non-linear evolution of ZNF stratified MRI by solving
the standard visco-resistive MHD equations,
\begin{eqnarray}
      \partial_t\rho +\nabla\cdt(\rho \V) & \!=\! & 0\,, \nonumber\\
      \partial_t(\rho\V) +\nabla\cdt
          [\rho\mathbf{vv}+p^{\star}\!\!-\mathbf{BB}] & \!=\! &
           2\rho\Omega\,( q \Omega x\,\xx - \zz\tms\V ) \nonumber
           - \rho\nabla\Phi(z) \zz \\ & & 
           + \rho\nu\nabla\cdt\left[\nabla\V +(\nabla\V)^{\top}
             -\nicefrac{2}{3}\nabla\cdt\V\,{\mathrm I}\right]\,, \nonumber\\
      \partial_t \B -\nabla\tms(\V\tms\B -\eta \nabla\tms\B) & \!=\! & 0\,,
      \nonumber
\end{eqnarray}
neglecting the effects of self-gravity and radiative
transport. Assuming an isothermal equation of state, the
stratification is thus fully defined by a static potential
$\Phi(z)$. The total pressure $p^{\star} = p + \nicefrac{1}{2}\B^2$,
all other symbols have their usual meanings. In the following, we will
mainly refer to the Lagrangian velocity $\U=\V-q\Omega x \yy$. We
focus on the case with box dimensions of $H\times\pi H\times 6H$ at a
resolution of $96/H$ and chose a fiducial value of
$\Pm=\nu/\eta=2$. Applying the spectral analysis described in
Sec.~3.1.2 of \cite{2007A&A...476.1123F}, we check that our choice of
${\rm Re}=H^2\Omega\,\nu^{-1}=6250$ is reasonably resolved, i.e., that
the estimated dissipation rate due to the numerical truncation error
is lower than the one given by the explicit dissipation.

For our simulations, we use the second order Godunov code NIRVANA
\citep{2004JCoPh.196..393Z} which has been extended and tested for the
shearing box formalism \citep{2007CoPhC.176..652G}. Since MRI
turbulence is inherently sub-Alfv{\'e}nic and transonic, accurate
treatment of the underlying MHD waves is mandatory. To improve the
effective resolution of our code at discontinuities, we implemented
the HLLD Riemann solver of \cite{2005JCoPh.208..315M}.

\begin{figure}
  \center\includegraphics[width=\columnwidth]{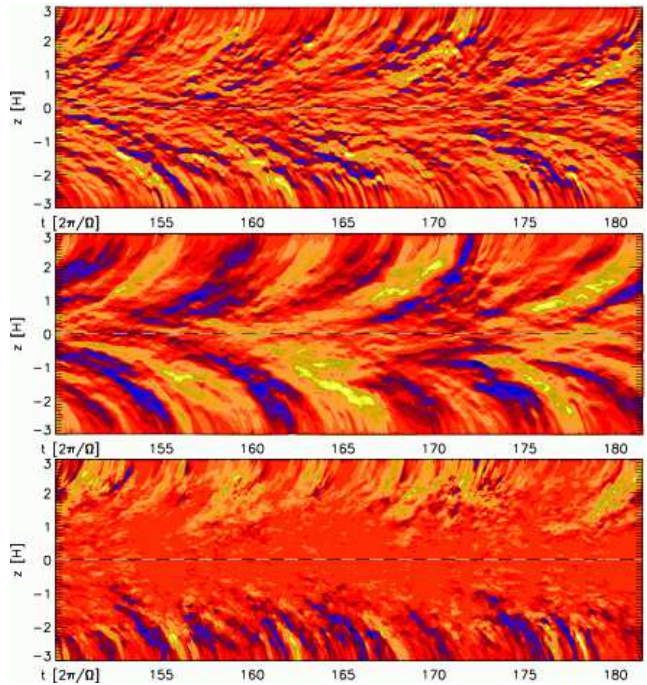}
  \caption{Space time evolution of the horizontally averaged radial
    field $\bar{B}_x(z,t)$, azimuthal field $\bar{B}_y(z,t)$, and
    magnetic $\alpha$~effect (cf. Sec.~\ref{sec:torsality}). The
    colour coding is normalised by the vertical rms amplitude at any
    given time to remove the stochastic fluctuations in the overall
    field strength and highlight the coherent pattern.}
  \label{fig:bfly_dns}
\end{figure}

Our vertical boundary conditions (BCs) are of the outflow type. This
allows the field to escape the box rather than pile-up and overshoot
as in the case of periodic vertical BCs
\citep[e.g.][]{1996ApJ...463..656S,2009arXiv0909.1570D}. For reasons
of robustness, the radial and azimuthal field components are set to
zero at these boundaries, which notably leads to the formation of a
disc ``wind'' as seen in panel (b) of
Fig.~\ref{fig:tfield}. Zero-gradient BCs in these fields work equally
well but no wind is seen in this case \citep[also
cf.][]{2009ApJ...691L..49S}. Apart from this, the features we are
concerned with in this study are, however, largely independent of the
boundaries (and the vertical extent of the box) and have been observed
with various types of BCs: potential field
\citep{2008AN....329..725B}, outflow \citep{2000ApJ...534..398M}, and
solving a characteristic equation \citep{2009ApJ...691L..49S}.

One argument in favour of periodic boundaries is that they conserve
the azimuthal and radial flux and therefore do not ``pollute'' the
dynamo-generated field via an inward Poynting flux
$S=\mathbf{E}\times\mathbf{B}$. We have checked that for $S_z=B^2
v_z-\mathbf{B\!\cdot\!v}\,B_z$, the dominant contribution is from the
advective term, and thus the net Poynting flux is directed
outwards. To compensate for the mass loss through the open boundaries
and provide a stationary background with respect to the hydrostatic
equilibrium, we implemented a continuous mass supply similar to the
one suggested by \cite{2009A&A...498..335H}.

\section{Simulation results}

In the middle panel of Figure~\ref{fig:bfly_dns}, we plot the temporal
evolution of the mean toroidal field $\bar{B}_y(z,t)$, showing
characteristic cycles on timescales of roughly ten orbits. Such cycles
occur naturally in a stratified environment
\citep{1995ApJ...446..741B,1996ApJ...463..656S,2000ApJ...534..398M,%
2004ApJ...605L..45T,2009ApJ...691L..49S,2009ApJ...697.1269J,%
2009arXiv0909.1570D}. Similar cycles have been
observed in non-stratified boxes with sufficient vertical extent
\citep{2008A&A...488..451L}, but it is currently unclear whether these
phenomena are related.

The parity of the field is not well defined and changes chaotically
from dipole to quadrupole symmetry, with the former prevailing. The
intermittent parity suggests that the underlying mechanism operates in
a highly non-linear and probably chaotic regime. The most striking
feature in Fig.~\ref{fig:bfly_dns}, however, are accelerated
"updraughts", reminiscent of the solar butterfly diagram.\footnote{We
caution the reader to bear in mind that this phenomenon might be
specific to the local box geometry, which is the scope of this
analysis.}

The evolution of the radial field is shown in the upper panel of
Fig.~\ref{fig:bfly_dns}, where the same upward motion is
visible. Unlike the toroidal field, the structures in the space time
diagram are much more filamentary \citep[also cf. Fig.~7
in][]{2009arXiv0909.1570D}. One is tempted to identify the cycle from
the blue and yellow lines, but this is misleading. When looking at a
slice at constant time $t$, these pronounced streaks partly exhibit a
bipolar structure in the \emph{vertical} direction rather than a
cyclic behaviour in time. At a given time, these streaks are somewhat
reminiscent of MRI channel modes. In fact, the first ones directly
emerge out of the initial linear growth phase. It remains open whether
such features survive at realistic magnetic Reynolds numbers,
$\Rm$. Ignoring the high contrast features, the same cycle as in $B_y$
becomes visible in red and orange colours.

Currently, there exist two scenarios which try to explain these
characteristic field patterns: on the one hand, it has been speculated
that the origin of these upward motions is buoyant rise due to
Parker-unstable toroidal fields \citep[see
e.g.][]{2000ApJ...534..398M}.  Alternatively,
\cite{1995ApJ...446..741B} have suggested the presence of an
$\alpha\Omega$~type mean-field dynamo, which can produce patterns that
travel away from the midplane if the $\alpha_{yy}$ component of the
dynamo tensor is negative.

In a recent analysis, \cite{2009arXiv0909.2003S} show that, several
scale heights $H$ away from the midplane, the undulatory Parker
instability is responsible for the upward magnetic motions. Near the
midplane, however, they find the flow to be buoyantly stable.

In the subsequent analysis, we will show by means of a simple
experiment that the field patterns near the midplane are effected by a
turbulent electromotive force (EMF). Only beyond $\sim 1.5\,H$ does
the bulk motion of the fluid become the dominant transport process.

\subsection{Mean-field description}

For a differentially rotating medium, the mean-field induction
equation of resistive MHD in the local Hill system reads
\begin{equation}
  \partial_t \mB = \nabla \times \left[\ \mU\tms\mB
    +\overline{\U'\tms \B'} + (q\Omega x\yy)\tms\mB 
    - \eta\nabla\tms\mB\ \right] \,, \label{eq:MF_ind}
\end{equation}
with $\eta$ the molecular diffusivity, and where overbars denote
horizontal averages, and primes denote the corresponding fluctuations
in the fluid velocity $\U$, and magnetic field $\B$. The shear
parameter $q$ takes the value of $-3/2$ for the case of a Keplerian
rotation profile.

 We want to stress that this formulation does not make any closure
assumptions, which means Eq.~(\ref{eq:MF_ind}) is
exact.\footnote{Given the cross terms $\overline{\U'\times\mB}$ and
$\overline{\mU\times\B'}$ vanish, which is the case if the chosen
averages comply with the Reynolds rules (idempotence of the averaging
operator, vanishing means of fluctuations) as is trivially fulfilled
for arithmetic averages.} The effect of the turbulence on the mean
field is expressed by the correlation $\EMF=\overline{\U'\tms \B'}$,
i.e., the mean of the cross product of the fluctuating velocity and
magnetic field.

\begin{figure}
  \center\includegraphics[width=\columnwidth]{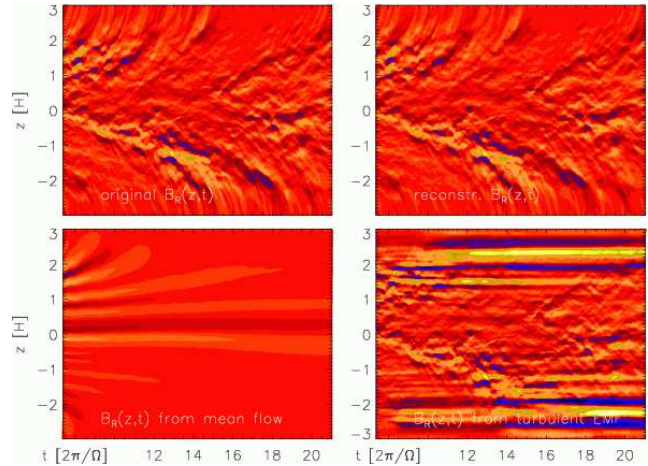}
  \caption{Reconstruction of $\bar{B}_x(z,t)$ from
    Eq.~(\ref{eq:MF_ind}), using space time profiles of $\EMF(z,t)$
    and $\bar{u}_z(z,t)$ stored from the simulation (upper panels,
    left: original, right: reconstructed). By individually
     discarding the two induction terms, we investigate
    the origin of the upward motions (lower panels, left: effect of
    mean flow, right: effect of turbulent EMF). Near the midplane, the
    field patterns are evidently due to the EMF. Buoyancy becomes
    dominant above $z\sim 1.5 H$.}
  \label{fig:recon}
\end{figure}

In the upper two panels of Fig.~\ref{fig:recon}, we demonstrate that
the collapsed profiles $\EMF_y(z,t)$ and $\bar{u}_z(z,t)$ contain all
the information to restore the mean radial field via
Eq.~(\ref{eq:MF_ind}), more specifically
\begin{equation}
  \partial_t\,\bar{B}_x = \partial_z\,(\bar{u}_z(z,t)\bar{B}_x +
  \EMF_y(z,t) - \eta \partial_z \bar{B}_x)\,.\label{eq:MF_ind_x}
\end{equation}
This implies that (in our 1D approach) only $\partial_z \EMF_y$ is
relevant to sustain $\bar{B}_x$ -- note that on the contrary
\cite{2009arXiv0909.1570D} find indications that $\partial_y \EMF_z$
may also play a role in replenishing the radial field. This does,
however, not seem to have an effect on horizontally averaged
quantities.

 Since $\partial_t \bar{B}_y$ is dominated by stretching of the radial
field via the shear term, we focus on the reconstruction of
$\bar{B}_x$, where only the first two terms in Eq.~(\ref{eq:MF_ind_x})
act as sources. By individually discarding these terms, we examine
their effect on the observed field pattern. If the turbulent EMF is
omitted (third panel in Fig.~\ref{fig:recon}), the initial field
decays. Field advection is only apparent away from the midplane, which
is consistent with the results of \cite{2009arXiv0909.2003S}. If, on
the other hand, we suppress the mean fluid motion $\mU$ and only
consider $\overline{\U'\tms \B'}$ (fourth panel in
Fig.~\ref{fig:recon}), we can still accurately reconstruct the rising
field structures near the midplane.

Away from the midplane, this approach is, of course, inconsistent, and
we would have to advect $\EMF(z,t)$ with $-\mU$ to match it with $\mB$
in a Lagrangian sense. The resulting grooves are again reminiscent of
MRI channel modes -- which are in fact valid solutions to the 1D
mean-field equation. In this respect, the scale separation is blurred
as one would naturally attribute such a small-scale effect to the
fluctuating field.


\section{Towards a possible accretion disc dynamo}

In the following, we set out to describe the cyclic behaviour in a
more generalised way, i.e., without a need for the time-dependent
electromotive forces from the direct simulations. To do so, we need to
apply a closure model to Eq.~(\ref{eq:MF_ind}). This is typically done
in the form of constant parameters which express the turbulent EMF in
terms of the mean field and its gradients. The introduced abstraction
is analogous to the $\alpha$~viscosity for the accretion stresses and
subsumes the effects of unresolved scales. Ultimately, the
universality of these parameters (or rather, their scaling with
respect to the relevant dimensionless numbers) needs to be probed. As
a first approach, we here apply a formulation which only retains
vertical derivatives \citep{2005AN....326..787B}:

\begin{equation}
  \EMF_i = \alpha_{ij}\,\bar{B}_j 
         - \tilde{\eta}_{ij}\,\varepsilon_{jkl}\,\partial_k \bar{B}_l\,,
  \qquad i,j \in \left\{x,y\right\}, k=z\,.
  \label{eq:closure}
\end{equation}

If we substitute this closure into the mean-field induction equation
(\ref{eq:MF_ind}), we obtain the standard $\alpha\Omega$~dynamo model,
where the diagonal elements of the $\alpha$~tensor give rise to a
feedback loop enabling exponential field amplification. In more
detail, $\alpha_{yy}$ describes the generation of poloidal field from
toroidal field via $\EMF_y=\alpha_{yy}\bar{B}_y$. Because of the
dominant shear term in the azimuthal field equation, $\alpha_{xx}$ is
usually sub-dominant for the operation of an $\alpha\Omega$ type
dynamo. Note, however, that the generation of azimuthal fields,
of vertical wave number $k_z$, is regulated by a term
$(\alpha_{xx}\,k_z + q\Omega)$, such that the radial $\alpha$~effect
can provide a saturation mechanism. We remark that this is potentially
interesting in view of the shear rate-dependence of the
Maxwell-to-Reynolds stress ratio \citep{2006MNRAS.372..183P}.

Neglecting the second term in Eq.~(\ref{eq:closure}), a simple
approximation to $\alpha_{yy}$ can be made by measuring the
correlation between $\bar{B}_y$ and $\EMF_y$. This has now been done
by various authors \citep{1995ApJ...446..741B,2001A&A...378..668Z,%
2002GApFD..96..319B,2009arXiv0909.1570D}, who find a negative
(positive) value for $\alpha$ in the top (bottom) half of the
box. \cite{1995ApJ...446..741B} remark that this effect has the wrong
sign with respect to what would be expected from quasi-linear theory
in the case of stratified rotating turbulence \citep[hereafter
RK93]{1993A&A...269..581R}. This has later been explained in terms of
magnetic buoyancy \citep{1998tbha.conf...61B}, and the idea was
subsequently confirmed by \cite{2000A&A...362..756R}. In the course of
the following analysis, we further elucidate this discrepancy by
showing that there exist two distinct \emph{indirect} dynamo
mechanisms.

\subsection{Kinetic and magnetic torsality} \label{sec:torsality}

Quasi-linear theory \citep{1980mfmd.book.....K} states that for
isotropic homogeneous turbulence there exists a kinematic
$\alpha$~effect
\begin{equation}
\aK=-\frac{1}{3}\tau_{\rm c}\,
                  \overline{\U'\cdot {\rm curl}(\U')}\,.\label{eq:akin}
\end{equation}
This term is derived from a closure in the induction equation to
obtain an approximation for $\partial_t \EMF^{\rm
kin}=\overline{\U'\tms\partial_t\B'}$, and describes the leading-order
effect of an imposed helical velocity field. As such, the kinematic
$\alpha$ neglects any feedback due to the magnetic field
itself. Considering the effect of the Lorentz force in the momentum
equation, \cite*{1976JFM....77..321P} derive a similar term for
$\partial_t \EMF^{\rm mag}= - \overline{\B'\tms \partial_t\U'}$, which
leads to an analogous magnetic $\alpha$~effect proportional to the
current helicity density $h_{\rm mag}=\B'\cdot {\rm curl}(\B')/\mu_0$
of the small-scale field,
\begin{equation}
\aM= \frac{1}{3}\tau_{\rm c}\,
                  \overline{\U'_{\rm A}\cdot {\rm curl}(\U'_{\rm
                  A})}\,,\label{eq:amag}
\end{equation}
where $\U'_{\rm A}=\B'/\!\sqrt{\mu_0\,\varrho}$ is the fluctuation
Alfv{\'e}n velocity. Approaching equipartition field strength, both
terms are equally important and add-up to an effective
$\alpha=\aK+\aM$. It is now a common notion in dynamo theory that
$\aK$ takes the role of the driver, while $\aM$ describes a non-linear
response, building up gradually and ultimately quenching the
kinematically imposed forcing. This process is a consequence of the
conservation of magnetic helicity and was termed \emph{dynamical}
quenching \citep{2002ApJ...579..359B}. However, see
\cite{2009arXiv0909.0721C} for an alternative method with a more
symmetric treatment of the momentum and induction equations. While
such an approach might be necessary in the presence of a magnetic
instability, we believe that the secondary large-scale dynamo effects
are well described by the dynamical quenching formalism.

\begin{figure}
  \center\includegraphics[width=0.95\columnwidth]{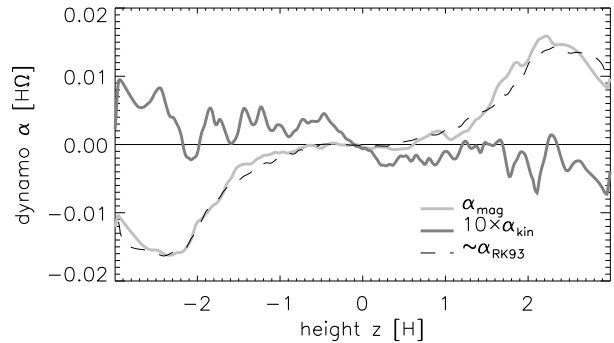}
  \caption{Helical $\alpha$~effect based on the current helicity
    (light grey) and kinetic helicity (dark grey). Both quantities are
    highly fluctuating in time but show a systematic average which
    follows the shape (dashed line) of the effect predicted for
    stratified, rotating turbulence by RK93.}
  \label{fig:heli}
\end{figure}

In Fig.~\ref{fig:heli}, we plot time averaged profiles for the two
quantities defined in Eqs.~(\ref{eq:akin}) and (\ref{eq:amag}) -- also
see the lower panel of Fig.~\ref{fig:bfly_dns}. Both torsalities are
dominated by strong rms fluctuations but show a systematic residuum if
averaged over times long compared to the cycle period. The magnetic
term dominates and strikingly shows exactly the shape that is expected
for stratified rotating turbulence \citep{1993A&A...269..581R}:

\begin{equation}
  \alpha_{\rm RK93} = -\tau_{\rm c}^2 \Omega\,\mn{u'^2} \left(\
  \Psi^{\rho}\,\nabla\log\rho \ +\ \Psi^{u}\,\nabla\log u' \
  \right)\,, \label{eq:alpha_strat}
\end{equation}
with rotational quenching functions $\Psi^{\rho}(\Co)$ and
$\Psi^{u}(\Co)$, weakly depending on the Coriolis number
$\Co=2\tau_{\rm c}\Omega$ of the flow.\footnote{For fast rotation,
turbulence becomes anisotropic along the axis of rotation. This is
however a minor effect in disc systems where $\Co< 1$.} Because of the
excellent agreement, we attribute this positive $\alpha$~effect to a
Parker-type instability. While we find the sign of $\aM$ to be robust
in all our simulations, the sign of $\aK$ curiously shows no
preference for the aligned or anti-aligned state.

To fix the free parameter in Eqs.~(\ref{eq:akin}) and (\ref{eq:amag}),
we roughly estimate $\tau_{\rm c}$ from a fit to the quasi-linear
approximation $\eta_{\rm t} \simeq \tau_{\rm c}/3\,\U'^2$ -- cf. panel
(c) of Fig.~\ref{fig:tfield} -- from which we infer a correlation time
$\tau_{\rm c}\simeq 0.2\Omega^{-1}$, corresponding to $\Co=0.4$. To
explain the observed amplitude in $\aM$, one however has to assume a
coherence time $\tau_{\rm c}^{\rm P}= 2.2\,\Omega^{-1}$ for the effect
in Eq.~(\ref{eq:alpha_strat}), which is enhanced by a factor of
$\sim10$ compared to the local coherence time. If this effect is
caused by a Parker dynamo \citep{1992ApJ...401..137P} operating above
$z \sim 1.5H$, which is consistent with the analysis of
\cite{2009arXiv0909.2003S}, the existence of a longer global turnover
time scale should however not be surprising.

Our amplitude of $\aM\!\sim\!0.02H\Omega$ is comparable to the value
measured by \cite{2008A&A...490..501J} for a strongly magnetised
accretion disc. Moreover, \citeauthor{2008A&A...490..501J} find
indications for a \emph{fast} dynamo, i.e., an increased effect for
higher $\Rm$.  This would imply that current simulations possibly
underestimate the saturation level of the turbulence. Finally, our
value for $\tau_{\rm c}^{\rm P}$ is also very similar to the one
estimated for zonal flows by \cite{2009ApJ...697.1269J}.

\begin{figure}
  \center\includegraphics[width=0.95\columnwidth]{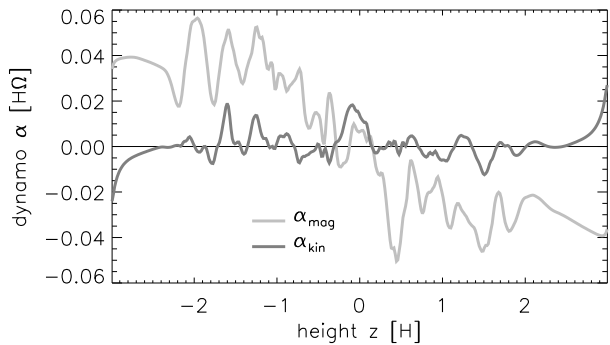}
  \caption{Same as Fig.~\ref{fig:heli}, but for a short time interval
    during the linear growth phase at the beginning of the
    simulation. Note the opposite sign of $\aM$ effected
    from MRI modes.}
  \label{fig:heli_MRI}
\end{figure}

The characteristic shape of the $\alpha$~profile in
Fig.~\ref{fig:heli} can be seen as an indication that the effect is
not caused by the MRI in an unintermediate way, but simply via the
stratified turbulence it creates. This notion finds further support
when looking at the initial linear growth stage of the MRI (see
Fig.~\ref{fig:heli_MRI}), where we observe a strong $\aM$ with the
opposite sign. Moreover, the distinct peaks in the profile can clearly
be identified with the MRI channels. Note that the tendency of these
features to migrate away from the midplane is consistent with the
negative sign of their magnetic torsality.  Figures~\ref{fig:heli} and
\ref{fig:heli_MRI} taken together support the conjecture that there
are \emph{direct} and \emph{indirect} dynamo mechanisms operating in
stratified MRI turbulence \citep[cf.][]{2004Ap&SS.292..395B}.

\subsection{Test field method}

In the past years, the so-called test field (TF) method
\citep{2005AN....326..245S,2008A&A...482..739B} has been established
as a standard tool to measure turbulent dynamo effects. The method
solves, simultaneously to the actual MHD simulation, a set of $\nu$
additional induction equations

\begin{eqnarray}\label{eq:TF}
\lefteqn{  \partial_t \Tf'_{(\nu)}  =  \nabla \times [\ 
    \U'\tms\mTf_{(\nu)} + (\mU\!+\!q\Omega x\yy)\tms\Tf'_{(\nu)}\ } \nonumber\\
    && \quad\quad
    - \ \overline{\U'\tms\Tf'}\!_{(\nu)} + \U'\tms\Tf'_{(\nu)} 
    - \eta\nabla\tms\Tf'_{(\nu)}\;] \,.
\end{eqnarray}
for the TF fluctuations $\Tf'_{(\nu)}$. Imposing $\nu\!=\!4$ suitably
varying fields $\mTf_{(\nu)}$, one thus obtains $4\times2$ linearly
independent EMF components, which allows to directly solve for the
eight unknown tensor coefficients $\alpha_{ij}$ and
$\tilde{\eta}_{ij}$.

In a more graphic way, the tracer fields pick-up the linear response
of the prescribed field under the effect of the small-scale velocity
field $\U'$. They do not directly see the actual magnetic field of the
simulation. However, since the velocity field $\U'$ is subject to the
Lorentz force, the method is well capable to capture a magnetically
induced $\alpha$ effect. Following \cite{2005AN....326..787B}, we use
the Fourier modes with $k_z=k_1\equiv 2\pi/L_z$ which should grasp the
essential behaviour for the fields seen in Fig~\ref{fig:bfly_dns}. The
ideal way of determination would be to apply TFs according to a
Fourier series \citep{2008A&A...482..739B}. Due to the growing number
of equations, this sets high demands on the computing power and is
currently not feasible given the resolution required to resolve the
most unstable MRI modes.

\cite{2009MNRAS.tmp.1219H} have recently demonstrated that the test
field method is applicable beyond the kinematic regime. The authors,
however, conclude that care has to be taken in interpreting the
results if (what they call) a meso-scale dynamo is present. We
conjecture that such an effect might be existent in our simulations in
the form of localised MRI modes. Whether such a \emph{direct} dynamo
is in fact significant, will have to be checked by further
investigations.

Because of the non-linear terms in Eq.~(\ref{eq:TF}), the test field
method is prone to the exponential amplification of small-scale
features \citep{2009MNRAS.395L..48C}. This has first been perceived as
a source of noise in the determination of the kinematic dynamo
effect. As a solution, \cite{2008MNRAS.385L..15S} and
\cite{2009PhDT........99G} suggest to reset the test field
fluctuations in regular time intervals. Such an approach has
previously been used for the imposed field method
\citep{2002A&A...394..735O}. Reset time intervals between $0.5$ and
$8$ orbital periods have been tested, and the results are found to
depend weakly on the chosen value \citep[also cf. lower panel of
Fig.~6 in][]{2009MNRAS.tmp.1219H}.

\begin{figure}
  \center\includegraphics[width=\columnwidth]{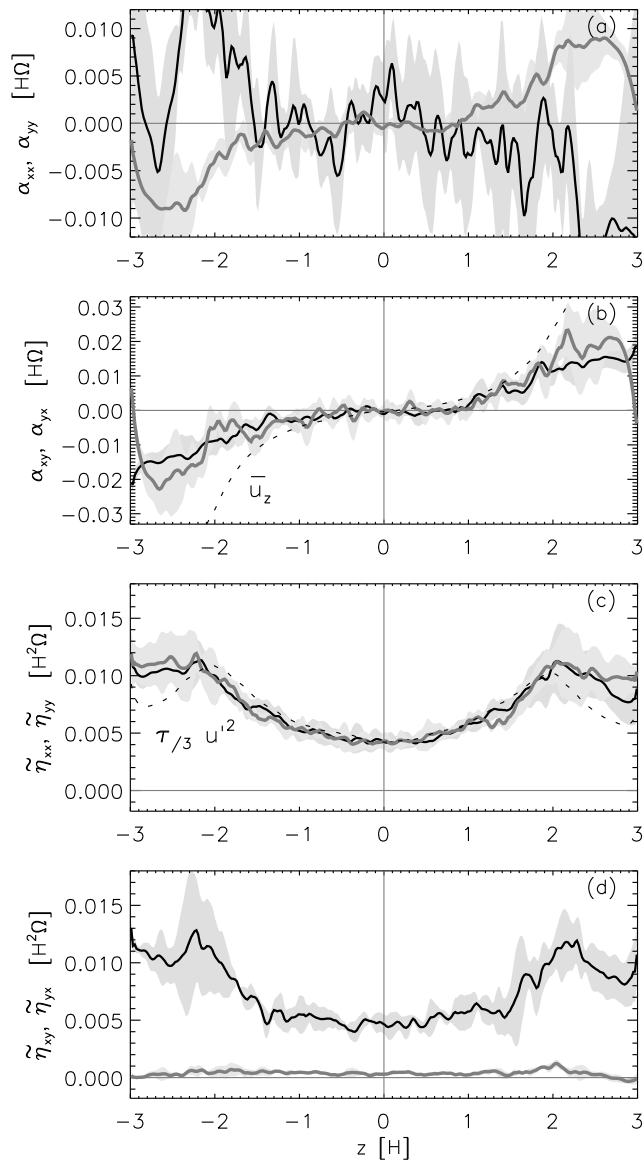}
  \caption{Dynamo coefficients obtained with the TF method. Ordinate
  labels refer to curves plotted in dark ($\alpha_{xx},\dots$) or
  light ($\alpha_{yy},\dots$) colours, respectively. Dashed lines show
  the mean vertical fluid motion, panel (b), and the quasi-linear
  estimate for the diffusion profile, panel(c).}
  \label{fig:tfield}
\end{figure}

\subsection{Tensor coefficients from the TF method}

The eight coefficients of the dynamo tensors are plotted as a function
of $z$ in Fig.~\ref{fig:tfield}. Most notably, the toroidal
$\alpha$~effect is found to be identical to $\aM$ in
Fig.~\ref{fig:heli}, i.e., positive (above the midplane) and follows
the analytical profile for stratified rotating turbulence of
RK93. Such profiles have also recently been observed in simulations of
buoyant galactic turbulence \citep{2008AN....329..619G}.
 
The shape of the diffusivity tensor $\tilde{\eta}$ (lower two panels)
and the relative strength of its components agree well with previous
results of \cite{2008AN....329..725B}, as do the off-diagonal elements
of the $\alpha$ tensor (second panel). We remark that
\cite*{2008arXiv0811.0542G} have found a negative kinematic $\alpha$
in both $\alpha_{xx}$ and $\alpha_{yy}$ for the case of stratified
Cartesian shear -- implying that MRI turbulence neither resembles
rotating, nor sheared turbulence. Moreover, unlike expected for a
kinematic dynamo, the radial and azimuthal components have opposite
signs. This might be indicative of a dynamo in a quenched state. It
therefore seems worthwhile to investigate whether the ratio of
$\alpha_{xx}$ and $\alpha_{yy}$ depends on the shear parameter $q$ in
a similar way as does the ratio of Maxwell and Reynolds stresses
\citep{2006MNRAS.372..183P}.

As already mentioned, the azimuthal $\alpha$~effect is positive
(negative) in the top (bottom) half of the box for $|z|>H$, but shows
the opposite sign near the midplane. This is in agreement with
\cite{2008AN....329..725B}, who found a very similar behaviour. The
\emph{negative} $\alpha$~effect near the mid plane can qualitatively
be explained by the buoyancy of small-scale flux tubes
\citep{1998tbha.conf...61B}.\footnote{Alternatively, this might simply
be the cumulative signature of persistent MRI channel modes
(cf. Fig.~\ref{fig:heli_MRI}), possibly seen as high-contrast features
in the uppermost panel of Fig.~\ref{fig:bfly_dns}.}  Even though the
effect is comparatively weak, the negative sign near the midplane
determines the overall dynamo mode which is required to explain the
observed direction of propagation \citep{1995ApJ...446..741B}.

Finally, we remark that earlier studies looking at the correlation
between $\bar{B}_y$ and $\EMF_y$ were potentially biased towards
regions of strong fields, which is consistent with a negative value
for $\alpha$ near the midplane.

\begin{figure}
  \center\includegraphics[width=\columnwidth]{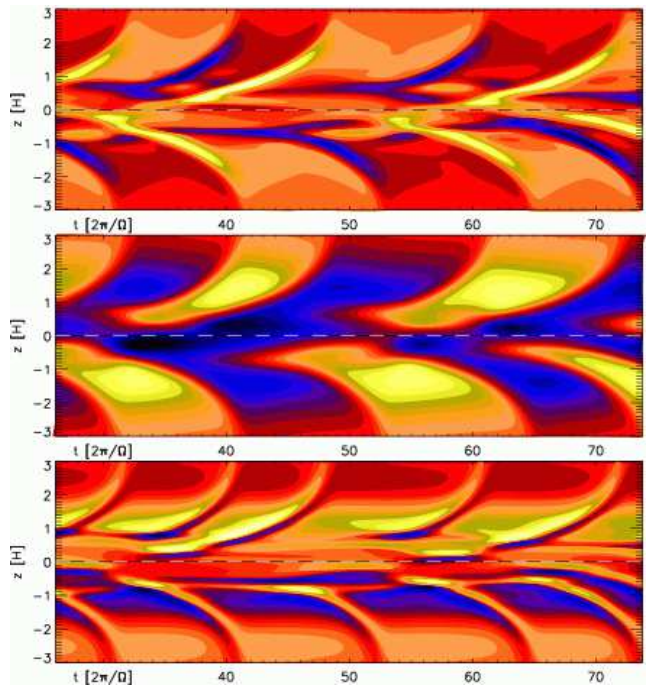}
  \caption{Dynamo patterns from a dynamically quenched 1D mean-field
    model. Quantities shown are $\bar{B}_x$ (top), $\bar{B}_y$
    (middle), and $\aM$ (bottom).}
    \label{fig:bfly_toy}
\end{figure}

\subsection{A dynamically quenched dynamo model}\label{sec:toy}

If we apply the inferred dynamo profiles to a simple one-dimensional
mean-field model, we can successfully reproduce the main features of
the butterfly diagram, as is shown in Fig.~\ref{fig:bfly_toy}. Note
that our approach recovers the asymmetry between the radial and
azimuthal field seen in Fig.~\ref{fig:bfly_dns}.

Unlike in earlier studies \citep{1997MNRAS.288L..29B,%
2002GApFD..96..319B}, and in addition to mean-field equations for
$\bar{B}_x(z,t)$ and $\bar{B}_y(z,t)$, we include an equation for
$\aM(z,t)$, the evolution of which is shown in the bottom panel of
Fig.~\ref{fig:bfly_toy}. The extra equation is motivated by the
dynamical quenching formalism, derived by \cite{2002ApJ...579..359B}
for the case of sheared turbulence. This formalism is founded on the
fundamental concept of magnetic helicity conservation. The dynamically
evolved value of $\aM(z,t)$ is, in turn, superimposed to the
prescribed kinematic $\aK(z)$, giving rise to genuinely non-linear
behaviour. We hereby closely follow the approach described in
Section~2.3 of \cite*{2009MNRAS.398.1414B} and implement their
Eq.~(17) assuming a simple advective helicity flux $\propto
\aM(z,t)\,\mU(z)$.

We here want to refrain from describing the related results in detail
as the system develops very complex behaviour for high magnetic
Reynolds numbers and a careful analysis is due. An extensive study on
the transition of the reduced system into chaotic behaviour seems
worthwhile.

We point out that the temporal evolution of $\aM$ already bears some
similarity with its counterpart in the direct simulations
(cf. Fig.~\ref{fig:bfly_dns}), and the overall parity
agrees. Moreover, there seems to exist a doubling of the cycle
frequency in this quantity, along with a distinct phase pattern with
one polarity prevailing. This intermittent pattern is likely related
to a phase shift between $\bar{B}_x$ and $\bar{B}_y$, which is a
hallmark of dynamo-generated fields.

The saturated amplitude of $\aM(z,t)$ is determined by the response to
the imposed $\aK(z)$. This is in agreement with the amplitudes of
$\aM$ and $\alpha_{xx}$ in Figs.~\ref{fig:heli} and \ref{fig:tfield},
respectively. Whether this is the correct direction of causality in
real MHD turbulence is as yet not a clear-cut question \citep[see][for
an alternative approach]{2004PhPl...11.3264B}.

\section{Conclusions}

As has now been found by a number of authors, vertical stratification
can alleviate concerns about non-convergent turbulent stresses in zero
net-flux shearing box simulations
\citep[see][]{2009arXiv0909.1570D}. Along these lines, the idea of
large-scale magnetic fields playing an important role in setting the
outer scale of the turbulence \citep{2007ApJ...668L..51P}, and hence
providing a meaningful amplitude for the viscous stress, has gained
new interest. Such fields are the natural outcome of a large-scale
dynamo, for which all requirements are met in a stratified shearing
box simulation. We propose that the saturation level of turbulent
stresses in such a scenario is intimately linked to the saturation of
these large-scale fields, making it mandatory to study the related
accretion disc dynamo.

In this paper, we have performed local simulations of stratified MRI
with zero vertical net-flux. Looking into the kinetic and current
helicities and probing the kinematic dynamo via the TF method, we have
identified a possible dynamo mechanism to explain the propagation of
mean magnetic fields away from the mid plane. Such an alternative
explanation is necessary because the flow is found to be stable to
Parker instability near the mid plane \citep{2009arXiv0909.2003S}, and
the pattern speed is independent of the bulk motion of the flow -- a
finding which strongly supports the interpretation in terms of a
dynamo wave.

Moreover, our analysis has brought forward a rather curious idea,
namely that there exist (at least) two distinct dynamo mechanisms --
one the immediate signature of MRI modes, and one the indirect effect
of the resulting turbulence in the presence of stratification. Such a
co-existence of a direct and indirect dynamo has already been
discussed by \cite{2004Ap&SS.292..395B}.

For the \emph{indirect} dynamo, there are two candidates: (i) a
classical Parker-type dynamo, i.e., ``cyclonic'' turbulence effected
by the Coriolis force \citep{1993A&A...269..581R}, and (ii) a
``buoyant'' dynamo caused by the Lorentz force. Such an effect has
first been predicted by \cite{1998tbha.conf...61B} and was derived
formally by \cite{2000A&A...362..756R}. While the former effect (with
positive $\aK$) is presumably dominant in the Parker-unstable halo,
the latter (with negative $\aK$) is likely to operate close to the
disc midplane.

Moreover, there is a remarkable resemblance to recent results by
\cite{2008A&A...490..501J}, who found a similar interplay between the
Parker instability and MRI for magnetically dominated accretion discs.
Our results similarly suggest to take up the ideas of a
self-regulatory dynamo cycle as proposed by \cite{1992MNRAS.259..604T}
almost two decades ago.

Whether this scenario is real, has to be checked by future studies. We
conjecture that the \emph{direct} dynamo should equally be seen in
non-stratified simulations as studied by
\cite{2008A&A...488..451L}.\footnote{Note, however, that even in the
absence of stratification, a dominant bipolar toroidal field might
still lead to a sufficient inhomogeneity in the turbulence to provide
a gradient in $\U'$.}

The central result of our analysis is the relevance of the current
helicity as a key indicator for magnetically induced dynamo action
\citep[cf. Sec.~3 of][]{2010AN....331..101B}. The picture is far from
being conclusive but the results are promising. Contrary to the
general scepticism towards kinematic dynamo theory in the context of
magnetic instabilities, the extension of mean-field theory with a
dynamical saturation mechanism (as demonstrated in Sec.~\ref{sec:toy}
by means of a simple 1D dynamo model) could well provide a framework
for understanding fully non-linear accretion disc dynamos.


\section*{Acknowledgements}
The author thanks Sebastien Fromang for discussions on improving the
numerical scheme and providing his spectral analysis tools. Richard
Nelson, Detlef Elstner and Axel Brandenburg are acknowledged for
helpful remarks on a draft version of this paper. I also thank the
anonymous referee for a careful report that helped to improve the
presentation of the paper. This work used the NIRVANA code version 3.3
developed by Udo Ziegler at the Astrophysical Institute Potsdam. All
computations were performed on the QMUL HPC facility, purchased under
the SRIF initiative.


\appendix

\bsp

\label{lastpage}

\end{document}